\documentclass[aps,twocolumn,superscriptaddress]{revtex4-2}

\usepackage[T1]{fontenc} 
\usepackage[utf8]{inputenc}
\usepackage{amsfonts,amsmath,amssymb}
\usepackage{mathtools}
\usepackage{nicefrac}
\usepackage{graphicx}
\usepackage{color}
\usepackage{soul} 
\usepackage{hyperref}
\usepackage{accents}
\usepackage{cancel}
\usepackage[normalem]{ulem}
\usepackage{dcolumn}
\usepackage{mathrsfs}
\usepackage{hyperref}
\usepackage{cases}
\usepackage{braket}
\usepackage{bbm}
\usepackage{bm}
\usepackage{mathptmx}
\usepackage{etoolbox}
 \usepackage{dutchcal}
\usepackage{tikz}
\usepackage{tikz-cd}
\usetikzlibrary{tikzmark}
\usepackage{capt-of}
\tikzcdset{scale cd/.style={every label/.append style={scale=#1},
    cells={nodes={scale=#1}}}}

\usepackage{braket}

\usepackage{parallel}
\usepackage{diagbox}
\usepackage{float}

\usepackage{adjustbox}

\def\tr{\mathrm{tr}}

\def\cK{{\cal K}}

\def\cH{\mathcal{H}}
\def\hcH{\hat{\mathcal{H}}}

\def\cB{{\cal B}}
\def\cE{{\cal E}}
\def\cO{{\mathcal O}}
\def\hcO{\hat{\mathcal{O}}}
\def\cS{{\mathcal S}}
\def\cC{{\cal C}}

\def\cD{{\mathscr D}}
\def\ci{{\mathcal i}}

\def\cU{\mathcal{U}}
\def\cB{\mathcal{B}}

\def\tU{\tilde{\mathcal{U}}}

\def\ks{\ket{s}}   
\def\bs{\bra{s}}

\def\bsp{\bra{s'}}
\def\khs{\ket{\hat{s}}}
\def\bhs{\bra{\hat{s}}}

\def\bhq{\bra{\hat{q}}}

\def\hs{\hat{s}}
\def\htw{\hat{t}}
\def\hq{\hat{q}}

\def\hX{\hat{X}}   
\def\hZ{\hat{Z}}
\def\hY{\hat{Y}}
\def\hQ{\hat{Q}}
\def\hP{\hat{P}}
\def\diid{\hat{\mathbbm{1}}}  
\def\iid{\mathbbm{1}}   
\def\hrho{\hat{\rho}}

\def\dag{^\dagger}

\def\bea{\begin{eqnarray}}
\def\eea{\end{eqnarray}}

\newcommand{\h}[1]{\hat{#1}}

\newlength{\dhatheight}

\begin{document}


\title{Quantum operations for Kramers-Wannier duality}



\author{Maaz Khan}
\email[]{maazkhan5507@gmail.com}
\affiliation{International Institute of Information Technology, Hyderabad (India)}

\author{Syed Anausha Bin Zakir Khan}
\email[]{anausha80@gmail.com}
\affiliation{Aligarh Muslim University, (India)}

\author{Arif Mohd}
\email[]{arif7de@gmail.com}
\affiliation{Aligarh Muslim University, (India)}


\date{\today}

\begin{abstract}
We study the Kramers-Wannier  duality for the transverse-field Ising lattice on a ring. A careful consideration of the ring boundary conditions shows that the duality has to be implemented with a proper treatment of different charge sectors of both the twisted and untwisted Ising and the dual-Ising Hilbert spaces.  We construct a superoperator that explicitly maps the Ising operators to the dual-Ising operators. The superoperator naturally acts on the tensor product of the Ising and the dual-Ising Hilbert space. We then show that the relation between our superoperator and the Kramers-Wannier duality operator that maps the Ising Hilbert space to the dual-Ising Hilbert space is naturally provided by quantum operations and the duality can be understood as a quantum operation that we construct. We provide the operator-sum representation for the Kramers-Wannier quantum operations and reproduce the well-known fusion rules. In addition to providing the quantum information perspective on the Kramers-Wannier duality,  our explicit protocol will also be useful in implementing the Kramers-Wannier duality on a quantum computer.   
\end{abstract}


\maketitle
\tableofcontents
\section{Introduction}
\label{sec:intro}
Kramers-Wannier duality of the two dimensional Ising model on a square lattice \cite{PhysRev.60.252} is a paradigmatic example of duality. While originally formulated for the statistical Ising model it can be converted via the transfer function formalism (see, for e.g., Ref.~\cite{Shankar_2017}) to a duality of the quantum Ising lattice in one dimension with a dual-Ising lattice that lives on the links of the original lattice. The duality also offers a theoretical laboratory in which many concepts like non-invertible symmetries, defect operators, entanglement entropy can be demonstrated, understood, and generalized 
(see, for e.g., the review article~\cite{shao2024whats}).

In this article our goal is to provide a quantum information perspective on the Kramers-Wannier duality  for the transverse-field Ising lattice on a ring.  We will show how the duality action on states can be understood as quantum operations and how the well known fusion rules can be derived through this perspective. Some articles in which similar ideas are discussed, implicitly or explicitly, are 
\cite{okada2024noninvertible, Ashkenazi_2022, Tantivasadakarn:2021vel, Tantivasadakarn:2022hgp, PRXQuantum.5.010338, Lootens_2023, lootens2023lowdepth}.

\begin{figure}
\includegraphics[scale=0.5]{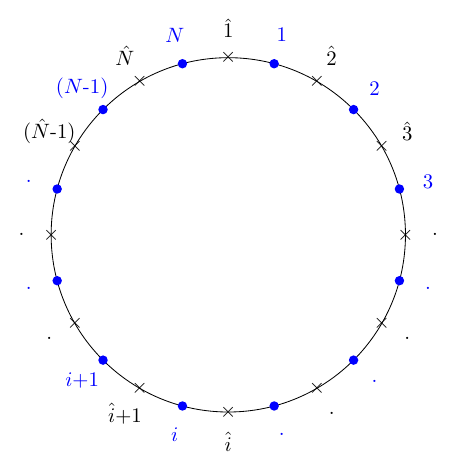}
\caption{Notation for the Ising (unhatted, blue) and the dual-Ising (hatted, black) spins on the lattice on the ring. The $i^{th}$ dual-Ising spin sits between the $(i-1)^{th}$ and the $i^{th}$ Ising spin.}
\label{fig:ising-ring}
\end{figure}

In order to frame our question sharply it  will be convenient to phrase the statement of the duality in terms of bond algebras (see Ref.~\cite{Cobanera_2011}). Let us denote the Ising operators/states with unhatted quantities and those of the dual-Ising with hatted quantities. 
The Ising bond algebra is the algebra generated by the operators  
$X_i$ and $Z_{i-1} Z_i$, where the index $i$ runs over the Ising spins 1 to N and $X, Y, Z$ are the usual Pauli operators. This algbera has global $\mathbb{Z}_2$ symmetry generated by the operator $Q= \prod \limits_{i=1}^N X_i$.
The dual bond algebra, i.e., the dual-Ising operator algebra, is generated by the operators $\hX_i$ and $\hZ_{i-1} \hZ_i$, and acts on the dual spins residing on the dual lattice
as shown in Fig.~\ref{fig:ising-ring}. The dual algebra also has a  global $\mathbb{Z}_2$ symmetry, sometimes called the quantum symmetry, generated by the operator $\hQ= \prod \limits_{i=1}^N \hX_i$.

The Kramers-Wannier duality, in terms of bond-algebras, can now be phrased as a map from the Ising Hilbert space $\cH$ to the dual-Ising Hilbert space
$\hat{\cH}$,
\begin{align}
\label{eq:mapD}
\cD : \cH \rightarrow \hat{\cH},
\end{align}
such that,
\begin{subequations}
\label{eq:cD}
\begin{align}
    \cD&X_i=\hat{Z}_i\hat{Z}_{i+1} \cD,\\ 
    \cD&Z_{i-1}Z_i=\hat{X}_i \cD.
\end{align}
\end{subequations}

We are however interested in the explicit mapping between the bond algebras themselves. It is clear that any operator that maps the bond algebras as in Eq.~\eqref{eq:cD}
must itself contain both the hatted (i.e., Ising) and unhatted (i.e., dual-Ising) operators. It thus naturally acts on the tensor product of the Ising and the dual-Ising Hilbert spaces, 
$\cH \otimes \h{\cH}$. Therefore the correct expression for the action of such an operator, now denoted by $\cU$, is 
\begin{subequations}
\label{eq:cU}
\begin{align}
    \cU&\left(X_i \otimes \diid\right) = \left(\iid \otimes \hat{Z}_i\hat{Z}_{i+1} \right) \cU,\\ 
    \cU&\left(Z_{i-1}Z_i \otimes \diid\right) = \left(\iid \otimes \hat{X}_i \right)\cU,
\end{align}
\end{subequations}
where the $\iid$ and $\diid$ are the identity operators on the Ising and the dual-Ising Hilbert spaces, respectively.

One of our objectives is to find an explicit expression of $\cU$. Another objective is to construct $\cD$ from this $\cU$. We will show that
$\cD$ acts on the space of states on the Ising Hilbert space $\cH$ (density matrices on $\cH$) as {\it quantum operations}. Not every quantum operation
admits a description in terms of state vectors (wavefunctions). Remarkably, $\cD$ does! We will show this and use this to obtain the well-known fusion rules for the
Kramers-Wannier duality.

This article is organized as follows. In Sec.~\ref{sec:KW}, we introduce the quantum Ising lattice on the ring and we discuss the boundary conditions and the charge/twist sectors. 
In Sec.~\ref{sec:ops} we construct the superoperator that implements the duality on the Ising bond algebra. In Sec.~\ref{sec:Qinf} we give a quick review of quantum operations.
In Sec.~\ref{sec:KWqops} we show how the Kramers-Wannier duality is realized as a quantum operation. Sec.~\ref{sec:discuss} concludes the paper with a summary and outlook.

\section{Ising-Ising duality on a ring}
\label{sec:KW} 
We have already stated the Kramers-Wannier duality in Eqns.~\eqref{eq:mapD} and ~\eqref{eq:cD}. Since the index $i$ runs over Ising spins from $1$ to $N$, for the complete specification of the duality operator $\cD$ we need to define the operators $Z_{0}$ and 
$\hZ_{N+1}$. 
 To that end, let us  analyze the action of $\cD$ on the charge $Q$. We have
\begin{align}
\cD Q &= \cD X_1 X_2 ... X_N \\
&= \hZ_1 \hZ_{N+1} \cD.
\end{align}
Now operating on $Q$ eigenkets with charge $\pm 1$,  $Q\ket{\pm 1} = \pm \ket{\pm 1} $ on both sides we get 
\begin{align}
\pm \cD \ket{\pm 1} = \hZ_1 \hZ_{N+1} \cD \ket{\pm 1},
\end{align}
which implies that $\hZ_{N+1} = \pm \hZ_1$, with the plus sign for the $Q=+1$ charge superselection sector of the Ising Hilbert space  and minus sign for the
$Q=-1$ superselection sector.
 It will be convenient to define a parameter $\htw = 0,1$, which we will soon interpret as ``twist'', and correspondingly consider the
instances of duality corresponding to two boundary conditions 
\begin{align}
\label{eq:dualtw}
\hZ_{N+1} = (-1)^{\htw} \hZ_1,
\end{align}
with $\htw=0$ as imposing the periodic boundary condition and $\htw = 1$ imposing the anti-periodic boundary condition on the dual-Ising system. We thus see that
under the Kramers-Wannier duality
charge $1$ states of the Ising Hilbert space are mapped to twist $0$  dual-Ising Hilbert space while charge $-1$ Ising states are mapped to the twist $1 $ dual-Ising Hilbert space. 
This can be summarized by the {\it fusion rule}
\begin{align}
\label{eq:fusionDQ}
\cD \times Q = (-1)^{\htw} \cD.
\end{align}
From a similar reasoning that led us to Eq.~\ref{eq:dualtw}, we have that
\begin{align}
\label{eq:dualt}
\hQ  \cD = \cD  Z_N  Z_0.
\end{align}
This time we operate on the charge $\pm 1$ eigenbras of $\hQ$ we deduce that  $Z_{0} = \pm Z_N$, with the plus sign for the  $\hQ=+1$ charge superselection sector of the dual-Ising Hilbert space and minus sign for the $\hQ=-1$ superselection sector. We define the Ising twist parameter $t = 0,1$ and consider the
instances of duality corresponding to two boundary conditions 
\begin{align}
\label{eq:tw}
Z_{0} = (-1)^{t} Z_N,
\end{align}
with  $t=0$ as imposing the periodic boundary condition and $t = 1$ imposing the anti-periodic boundary condition on the Ising system. As above, we have that under the Kramers-Wannier duality
charge $1$ states of the dual-Ising Hilbert space are mapped from the twist $0$ sector of Ising Hilbert space while charge $-1$ dual-Ising states are mapped from the twist $1$ Ising sector. 
This can be summarized by the {\it fusion rule}
\begin{align}
\label{eq:fusionQD}
\hQ \times \cD  = (-1)^{t} \cD.
\end{align}
We thus have the following checkerboard~\cite{Seiberg_2024, Li_2023, shao2024whats} for the Kramers-Wannier duality,

\begin{table}[H]
   
\begin{center}
\begin{tabular}{ |c|c|c| } 
\hline
 \backslashbox{Q}{t} & 0 & 1 \\
 \hline
 +1 &  A & B\\
 \hline
 -1 & C & D \\
 \hline
\end{tabular}
\qquad
\begin{tabular}{ |c|c|c| } 
\hline
 \backslashbox{$\hat{Q}$}{$\hat{t}$} & 0 & 1 \\
 \hline
 +1 &  A & C\\
 \hline
 -1 & B & D \\
 \hline
\end{tabular}
\end{center}
 \caption{Mapping of the charge-superselection sectors and the twisted Hilbert spaces by the Kramers-Wannier duality. }
    \label{tab:kw}
\end{table}

So far we have not mentioned a specific Ising Hamiltonian. The only requirement we have is that it be constructed out of the bond algebra. Although the specific Hamiltonian
is not needed for the purposes of this paper, it is nevertheless instructive to have an example to understand the effect of twists $t$ and $\htw$. For the transverse-field Ising Hamiltonian we have
\begin{align}
\label{eq:IsH}
H = - \lambda \sum_{i=1}^{N} Z_i Z_{i+1} -  \sum_{i=1}^{N} X_i.
\end{align}
Note that there are two theories here depending upon the boundary condition in Eq.~\ref{eq:tw}. These are called the twisted sectors. 
Explicitly, the Hamiltonian in each sector is labelled by the twist $t \in \{0,1\}$ as
\begin{align}
\label{eq:IsingH}
H^{\{t\}} &= - \lambda \left[ \sum_{i=1}^{N-1} Z_i Z_{i+1} - (-1)^t Z_N Z_1 \right] -  \sum_{i=1}^{N} X_i. 
\end{align}
 Now applying the duality transformation in Eq.~\ref{eq:cD} to the Ising Hamiltonian we get 
 \begin{align}
 \cD H = \hat{H} \cD,
 \end{align} 
where the dual-Ising Hamiltonian is 
\begin{align}
\label{eq:dwH}
\hat{H} = - \lambda  \sum_{i=1}^{N} \hX_i - \sum_{i=1}^{N} \hZ_i \hZ_{i+1}.
\end{align}
Just like the Ising twist sectors, here also we have two dual-Ising twist sectors that are governed by their own Hamiltonians labelled by $\htw \in \{0,1\}$ that we write explicitly as
\begin{align}
\label{eq:dwtH}
\hat{H}^{\{\htw\}} &= - \lambda  \sum_{i=1}^{N} \hX_i - \left[ \sum_{i=1}^{N-1} \hZ_i \hZ_{i+1} + (-1)^{\htw}\hZ_N \hZ_1 \right].
\end{align}

Note that the effect of the duality transformation on the theory is to invert the coupling $\lambda$, wherein lies the power of the duality to understand a strongly coupled theory in terms of 
a weakly coupled theory. We emphasize that it is {\it not} that that Kramers-Wannier duality is mapping the $t=0$ Ising sector to some particular $\htw$ dual-Ising sector.
The duality action is more nuanced. The fusion rules in Eqns.~\eqref{eq:fusionDQ} and ~\eqref{eq:fusionQD} relate
charge superselection sectors and twisted sectors across the duality as summarized in table~\ref{tab:kw}. Modern way to interpret this table, say for the entry labelled
as $C$, is that the charge $-1$ states of Ising are mapped to charge 1 states of dual-Ising with a defect. Similarly, the entry labelled $B$ says that the charge 1 states of the
defect Ising are mapped to the charge -1 states of dual-Ising. The twist, therefore, corresponds to the $\mathbb{Z}_2$-symmetry  defect ~\cite{shao2024whats}.

Next we want to construct the superoperator $\cU$ of Eq.~\eqref{eq:cU} which is an ingredient in our quantum operation perspective to be developed later.

\section{Construction of the superoperator $\cU$}
\label{sec:ops}
From Eq.~\eqref{eq:cU} the action of $\cU$ on  charges $Q$ and $\hQ$ can be obtained as
\begin{subequations}
\label{eq:cUQ}
\begin{align}
   \cU \left(Q \otimes \diid \right) &= (-1)^{\htw} \cU,\\ 
   (-1)^t \cU &= \left(\iid \otimes \hQ \right)\cU,
\end{align}
\end{subequations}
 where we have also made use of the boundary conditions of Eqns.~\eqref{eq:dualtw} and \eqref{eq:tw}. We will construct $\cU$ in two steps.
 
 We first define an operator $\cS$  that implements the duality transformation like Eq.~\ref{eq:cD} on a {\it single} Ising chain in the sense that
\begin{subequations}
\label{eq:cS}
\begin{align}
    \cS\hat{X}_i &=\hat{Z}_i\hat{Z}_{i+1} \cS, \\   
    \cS\hat{Z}_{i-1}\hat{Z}_i &=\hat{X}_i \cS.
\end{align}
\end{subequations}
The explicit expression for this operator (see also Ref.~\cite{shao2024whats}) is 
\begin{align}
\label{eq:single}
\cS=  \hP \left[ \prod_{i=1}^{N-1}\frac{(1+i\hat{X}_i)}{\sqrt{2}}\frac{(1+i\hat{Z}_i\hat{Z}_{i+1})}{\sqrt{2}} \right]\frac{(1+i\hat{X}_N)}{\sqrt{2}},
\end{align}
where the hatted projector $\hP$ is,
\begin{align}
\label{eq:hP}
\hP=\frac{1+ (-1)^{t} \hQ}{2}.
\end{align}
Next we define an operator $\cC$ that copies the Ising operators to the dual Ising operators just in the right way so that the duality is implemented when followed by $\cS$ in Eq.~\eqref{eq:single}.
In particular,
\begin{subequations}
\begin{align}
    \cC \left(X_i \otimes \diid\right) &= \left(\iid \otimes \h{X}_i\right)  \cC, \\
    \cC \left({Z}_{i-1}{Z}_i \otimes \diid\right) &=\left(\iid \otimes \hat{Z}_{i-1}\hat{Z}_i \right) \cC.
\end{align}
\end{subequations}
The explicit expression of the copying operator is
\begin{align}
\cC = \left[\prod_{i=1}^{N-1}\left(SWAP\right)_i\right] \frac{(1+iX_N\hX_N)}{\sqrt{2}}\frac{(1+iY_N\hY_N)}{\sqrt{2}} P,
\end{align}
where $(SWAP)_i$ is the operator that swaps between the $i^{th}$ Ising and dual-Ising spins,
\begin{align}
\label{eq:swap}
(SWAP)_i = \frac{1}{2} \left(1 + X_i\hX_i + Y_i\hY_i + Z_i\hZ_i \right),
\end{align} 
and $P$ is the projection,
\begin{align}
\label{eq:P}
P=\frac{1+ (-1)^{\htw} Q}{2}.
\end{align}

Finally, the superoperator that  implements the duality transformation on the Ising operators in the sense of Eq.~\ref{eq:cU} is
\begin{align}
\label{eq:superU}
\cU :=  \cS\cC . 
\end{align}
$\cU$ is the key operator of this paper. Due the presence of projectors in $\cS$ and $\cC$, the operator $\cU$ is non-invertible. 
The projectors are there to ensure that the algebra in Eq.~\eqref{eq:cUQ} is respected. 
 For later use we also define an operator $\tU = P \cU \hP $ 
  and record the following relations,
\begin{subequations}
\begin{align}
\cU \dag \cU &=  \frac{P}{2}, \\
\tU \dag \tU &= \frac{P \hP}{2}. \label{eq:UdagU}
\end{align}
\end{subequations}

 What is the relation between $\cU$ of Eq.~\eqref{eq:cU} and $\cD$ of Eq.~\eqref{eq:cD}? How do we get $\cD$ from $\cU$? We claim that the answer to these questions is obtained by viewing the Kramers-Wannier duality through the lens of quantum information theory.
More precisely, we will show that action of $\cD$ is that of a non trace-preserving quantum channel, called quantum operations in general, acting on
 the density matrices on the Ising Hilbert space $\cH$ and yielding density matrices on the dual-Ising Hilbert space $\hat{\cH}$.
In the next section, Sec.~\ref{sec:Qinf}, we quickly review the quantum operations before moving on to show how the Kramers-Wannier duality is framed in this language. 

\section{A quick review of Quantum Operations}
\label{sec:Qinf}
In this section we give a quick review, based on Ref.~\cite{nielsen2001quantum} of the definition of quantum operations and the equivalent form in which we will use it in the next sections.
For details we refer the reader to Ref.~\cite{nielsen2001quantum}.

A quantum operation $\cE$ is defined as a map  from the  density operators on  $\cH$ to the  density operators on
$\h{\cH}$
that satisfies the following three properties:
\begin{enumerate}
\item tr $ \left[ \cE (\rho)\right] \leq 1$
\item $\cE$ is a convex linear map on the set of density matrices on $\cH$.
\item $\cE$ is a completely positive map.
\end{enumerate}
When $ \tr\left[ \cE (\rho)\right] $ is strictly equal to $1$ quantum operation is referred to as quantum channel.
It can be proved (see, Ref.~\cite{nielsen2001quantum}) that a map $\cE$ satisfies the above three axioms if and only if it 
admits an operator sum representation,
\begin{align}
\cE(\rho) = \sum_i E_i \rho E_i \dag,
\end{align}
for some set of operators $\{E_i\}: \cH \rightarrow \h{\cH}$, such that the sum 
\begin{align}
\label{eq:comp}
\sum_i E_i \dag E_i \leq \mathbbm{1}.
\end{align}

 It is in this explicit 
form that we will express the quantum operation implementing the Kramers-Wannier duality.   

\section{Kramers-Wannier via Quantum Operations}
\label{sec:KWqops}
 Let us begin by fixing our notation. 
Both $X_i$ and $Z_i$ operators have eigenvalues of $\pm 1$. We will be working in the $Z_i$ eigenbasis and choose the eigenvectors of the $Z_i$ operators, $\forall i$  
to be $\ket{s}$, 
where $\forall i, s_i = 0, 1$ and $\ks = \otimes_i\ket{s_i}$ . We thus have, $Z_i\ks=(-1)^{s_i}\ket{s}$ and  
$X_i\ks=\ket{s_1}\otimes \ket{s_2} \otimes ...\ket{1-s_i}\otimes \ket{s_{i+1}}...\otimes \ket{s_N}$. For the dual-Ising we just put hats over everything.

We first construct the operator sum representation of the duality and then we will verify its validity by matching the correlation functions across the duality and then by matching with the known fusion rules. 

\subsection{Operator-Sum representation of Kramers-Wannier}
We will construct the quantum operation $\cD$ for the Kramers-Wannier duality via its {\it environment model}. See, for e.g., Ref.~\cite{nielsen2001quantum}, where it is shown that any 
quantum operation admits such a model. In our case the role of the environment is played by the dual-Ising Hilbert space. Technically speaking, we obtain the duality from its Stinespring dilation.
Let us first sketch the basic structure and then address the subtleties inherent to our specific application.
Our  quantum operation $\cD$ will be realized as  
shown in Fig.~\ref{fig:KWmaps}.
\begin{center}
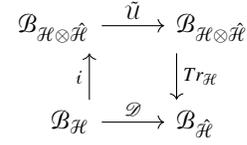

\begin{tikzcd}[scale cd=1]
 \mathllap{\cB_{\cH \otimes \h{\cH}}} \arrow[r, "\tU"]  & \mathrlap{\cB_{\cH \otimes \h{\cH}}} \arrow[d, "Tr_{\cH}"] \\
\mathllap{\cB_{\cH}} \arrow[u, "\ci"] \arrow[r, "\cD"] & \mathrlap{\cB_{\h{\cH}}}
\end{tikzcd}
\captionof{figure}{Obtaining the quantum operation $\cD(\rho)$ via its {\it environment model} (or its Stinespring dilation). The environment in this case is provided by the dual-Ising Hilbert space. 
}
\label{fig:KWmaps}
\end{center}

To unpack this diagram we first work out its decorated version as shown in Fig.~\ref{fig:KWkets}. Let us recall that for both the Ising and the dual-Ising we have two Hilbert spaces each, namely, the twisted and the untwisted. 
In each (un)twisted sector we also have the charge superselection sectors which, as discussed in Sec.~\ref{sec:KW}, give rise to the checkboard action of the duality as shown in table~\ref{tab:kw}.
In order to respect this checkerboard we have to introduce projections at all stages of our diagram. 
The map $\ci$ inserts the Ising density matrix $\rho$ into the tensor product Hilbert space $\cH \otimes \h{\cH}$ by tensoring its projection $P \rho P$
with an arbitrary but fixed projected pure state density matrix of dual-Ising $\hP \khs\!\bhs \hP $. We then apply the operator $\cU$ (note, it is not $\tU$) on the combined state 
and afterwords trace over the projected Ising spins. The result is the map $\cH \owns \rho \rightarrow \cD(\rho) \in \h{\cH}$ that can be obtained as follows: 
\begin{center}
\begin{tikzcd}[scale cd=1]
\mathllap{{P\rho P }\otimes {\hP \khs\bhs \hP} \arrow[r, "\cU"]}  & \mathrlap{\cU  (P\rho P \otimes {\hP \khs\bhs \hP}) \cU^{\dagger}  \arrow[d, "Tr_{P\cH}"]} \\
 \rho \arrow[u, "\ci"] \arrow[r, "\cD"] &\mathrlap{\cD(\rho)}.
\end{tikzcd}
\captionof{figure}{Decorated version of the duality mapping an Ising state into the dual-Ising Hilbert space. We call it decorated because of the various projections involved in all stages. 
$Tr_{P\cH}$ on the right arrow going down refers to the trace being taken by inserting the projection $P$.}
\label{fig:KWkets}
\end{center}

\begin{align}
\cD(\rho) &= \sum_{\{s\}} (\bs P) \cU  \left[P\rho P  \otimes \hP \khs\bhs \hP\right] \cU^{\dagger} (P\ks) \label{eq:cDrho} \\ 
&= \sum_{\{s\}} \bs (P \cU P \hP)  \left[\rho   \otimes  \khs\bhs\right] (P \hP \cU^{\dagger} P) \ks \\
&= \sum_{\{s\}} \bs \tU \left[\rho   \otimes  \khs\bhs\right] \tU \dag \ks \\
&= \sum_{\{s\}} \bs \tU \khs  \rho \bhs \tU{\dag} \ks \\
&= \sum_{\{s\}} D_{s} \rho D_{s}\dag, \label{eq:opsum}
\end{align}
where Kraus operators $D_{s} :=  \bs \tU \khs $, for each $s$ and fixed $\hs$, are maps: $\cH \rightarrow \h{\cH}$, and in the second step we have used the definition
$\tU = P \cU \hP $ 
and the fact that $\cU$ already contains a $\hP$ on the left and a $P$ on the right, see Eq.~\eqref{eq:superU}. Therefore, the projections can all be absorbed in $\tU$ and we 
have the exact unpacking of Fig.~\ref{fig:KWmaps} as shown in Fig.~\ref{fig:KWunp},
\begin{center}
\begin{tikzcd}[scale cd=1]
\mathllap{{\rho  }\otimes { \khs\bhs } \arrow[r, "\tU"]}  & \mathrlap{\tU  (\rho  \otimes {\khs\bhs}) \tU^{\dagger}  \arrow[d, "Tr_{\cH}"]} \\
 \rho \arrow[u, "\ci"] \arrow[r, "\cD"] &\mathrlap{\cD(\rho)}.
\end{tikzcd}
\captionof{figure}{Duality mapping an Ising state into the dual-Ising Hilbert space. All the projections are now absorbed in $\tU$.}
\label{fig:KWunp}
\end{center}
 
Next we need to verify the completeness relation in Eq.~\eqref{eq:comp}. We have 
\begin{align}
\sum_{s} D_{s}\dag D_{s} &= \sum_{s} \bhs \tU\dag \ks \bs \tU \khs \\
&= \bhs \tU{\dag}  \tU \khs \\
&= \frac{P}{4}\\
&< \iid,
\end{align} 
where in the second step we have used Eq.~\eqref{eq:UdagU}, and the last step follows because $P$ is a projector (see Eq.~\eqref{eq:P}). Therefore, $\cD(\rho)$ is a bona fide trace non-preserving
 quantum operation with its operator sum representation given in Eq.~\eqref{eq:opsum}. 
 Let us mention here that the triple $\pmb{\langle} \hcH, \tU, \khs\!\bhs \pmb{\rangle}$ is a Stinespring dilation of the quantum operation $\cD(\rho)$ (see, for e.g., Ref.~\cite{Heinosaari_Ziman_2011}). 
 
 In what sense is $\cD$  implementing the Kramers-Wannier duality on state vectors? It is true that our $\tU$ (or $\cU$) is constructed precisely to implement the duality on operators.
  But how does this imply that
 $\cD$ does the same for states? We address these questions now, first answering it for density matrices and then for state vectors. Finally we reproduce the fusion rules for the duality.

 \subsection{Why is $\cD$ the Kramers-Wannier duality?}
The basic tenet of any duality is that it  preserves the correlation functions, up to possibly an overall state and operator independent normalization. 
The question is therefore if the quantum operation $\cD$ preserves 
correlation functions in the sense that
\begin{align}
\label{eq:KWcor}
Tr_{\cH}\left({\cO \rho}\right) \propto Tr_{\hcH} \left(\hcO \hrho\right),
\end{align}
where, $\hrho = \cD(\rho)$, and where, $\cO$ and $\hcO$ are operators of the Ising and the dual-Ising bond algebras, respectively, such that 
\begin{align}
  \tU&\left(\cO \otimes \diid\right) = \left(\iid \otimes \hcO \right) \tU.
\end{align}

It is easy to show that 
\begin{align}
Tr_{\hcH} &\left(\hcO \cD(\rho)\right) =  \frac{1}{4} Tr_{\cH} \left(\cO P\rho P\right).
\end{align}
However, due to the charge superselection in force we have $P \rho P = \rho$. In our definition
of insertion $\ci$ in Fig.~\ref{fig:KWkets} we had inserted $\rho$ in its projected form. We could
have alternatively restricted to the density matrices compatible with the charge superselection,
in which case we have the equality of correlators up to an irrelevant overall constant. Thus our
quantum operation $\cD$ does implement the Kramers-Wannier duality with the understanding of the map of  
charge superselection and twisted sectors as shown in table~\ref{tab:kw}.

In order to understand the duality in terms of vector states as stated in Eqns.~\eqref{eq:mapD} and \eqref{eq:cD} we
calculate the action of $\cD$ on a general basis element of the form $\ks\!\bsp$. A straightforward calculation gives,
\begin{align}
\cD({\ks\!\bsp}) = 2 (-1)^{(t + \htw)s_N} \cS\ket{\hs = s} \bra{\hs = s'}\cS\dag (-1)^{(t+\htw)s'_N},
\end{align}
from which we may read off the action on the basis of vector states.
Bearing a slight abuse in notation and using the same $\cD$ to denote the action on state vectors,
\begin{align}
\cD \ks &=  \sqrt{2}\, (-1)^{(t + \htw)s_N} \cS\ket{\hs = s} \\
&=\sqrt{2}\, \cS \, \hZ_N^{(t + \htw)} \ket{\hs = s},
\end{align}
from which we can calculate the matrix elements of $\cD: \cH \rightarrow \hcH$. For $\cD_{s \hq } = \bhq \cD \ks$, we get
\begin{align}
\cD_{s \hq } = \frac{1}{2^{N/2}} \, e^{i (N-2t)\frac{\pi}{4}} \, (-1)^{\sum \limits_{i=1}^{N}\hq_i (s_i - s_{i-1}) + s_N \htw},
\end{align}
which matches with the one quoted in Ref.~\cite{Li_2023} from Ref.~\cite{Aasen_2016} up to an irrelevant phase. Using these matrix
elements  the fusion rules in Eqns.~\eqref{eq:fusionDQ} and ~\eqref{eq:fusionQD} are easily seen to be satisfied. Furthermore,
the $\cD\dag \cD$ fusion rule can be obtained to be
\begin{align}
\label{eq:fusionDD}
\cD \dag \times \cD = 1 + (-1)^{\htw}Q.
\end{align} 

Thus we see that  our quantum operation $\cD$ preserves the correlation functions and reproduces the fusion rules for Ising-Ising duality. We have therefore reached our stated goal in this work to implement the Kramers-Wannier duality via quantum operations.  

\section{Discussion}
\label{sec:discuss}
In this paper we have provided a quantum information perspective on the Kramers-Wannier duality of the  Ising lattice on a ring. We showed that the duality acts on Ising states (density matrices) and on the Ising bond algebra via quantum operations. We gave an explicit construction of this quantum operation and reproduced through this perspective the duality action on the vector states and the well-known fusion rules in the literature \cite{Li_2023,shao2024whats}. Due to the charge superselection sectors and the inclusion of twisted Hilbert spaces the duality is not formulated on a fixed Hilbert space. This 
is reflected in the non trace-preserving nature of our quantum operation implementing the duality. Nevertheless the duality is still useful because it maps the Ising correlators to the dual-Ising correlators as we have shown. This has an effect of inverting the coupling and therefore mapping a strongly coupled system to a weakly coupled system exemplifying the UV/IR nature of dualities. If one insists on formulating the duality on a fixed Hilbert space then one would have to restrict to the individual charge superselection sectors~\cite{Radicevic:2018okd}, in which case we would have a trace preserving quantum channel. 

Yet another way in which Kramers-Wannier duality can be derived is by gauging the $\mathbb{Z}_2$ charge  of the Ising model \cite{Aksoy:2023hve, RevModPhys.51.659, RevModPhys.52.453}. It will be interesting to see how the gauging approach can be seen from the quantum operations perspective (see, for e.g., Ref.~\cite{Ashkenazi_2022}) and how it is related to our approach here which does not involve gauge fields.  This may also pave the way to thinking about `t Hooft anomalies from the point of view of quantum information theory. Let us elaborate on this a little. In either the original argument of `t Hooft~\cite{tHooft:1979rat} or that of Callan and Harvey~\cite{CALLAN1985427}, some ancillary degrees of freedom are coupled to the original system with a global symmetry. In the case of `t Hooft these degrees of freedom are the background gauge field  for the global symmetry and chiral fermions coupled to this gauge field that cancel the gauge anomaly, while in the case of Callan-Harvey the ancillary degrees of freedom live in higher dimensions and cancel the gauge anomaly by the anomaly-inflow mechanism. In both cases, in the end the ancillary degrees of freedom are decoupled (traced out) 
and one deduces that the `t Hooft anomaly should remain the same at any energy scale \footnote{In the original argument of `t Hooft the ancillary degrees of freedom are not  traced out but their coupling with the original theory is set to zero.}. It is not suprising that this procedure sounds a lot like quantum operations. After all, quantum operations are the most general transformations, including measurements, that can be applied to a system.  

Very recently Ref.~\cite{okada2024noninvertible} appeared that discusses linear Ising chain (as a demonstration of a more general observation of these authors) from the point of view of topological defect operators. The main thesis of Ref.~\cite{okada2024noninvertible} is that non-invertible symmetries (which, in the Ising case, correspond to inserting the Kramers-Wannier duality defect) act on local operators by quantum operations. Our work can be thought of as providing a detailed operator (instead of the defect) demonstration of Ref.~\cite{okada2024noninvertible} for the Ising model on the ring. The Stinespring triple $(V, \pi, \cK)$ that Ref.~\cite{okada2024noninvertible} finds for the action of non-invertible symmetries on local operators is, in our language \footnote{A minor technicality is that Ref.~\cite{okada2024noninvertible} inserts two defects. For a strict translation to our language we will have to concatenate two quantum operations: Ising to dual-Ising, and then back to Ising. It would be interesting to fill up the details. } 
$\left(V=\tU \dag\ks,\, \pi(\cO) =  \cO \otimes \diid, \,\cK = \cH \otimes \hcH \right)$.  
 The generality of the beautiful argument of Ref.~\cite{okada2024noninvertible} suggests that the quantum operations picture should apply to all dualities, or at least to those that can be obtained by insertions of defects corresponding to non-invertible symmetries. It will be interesting to construct an explicit argument to this effect in the language of quantum operations. It will also be interesting
to understand the structure theory of quantum operations as to which quantum operations correspond to non-invertible symmetries. 

Finally, as mentioned in the abstract,  the  explicit operators we have provided in this paper makes it possible to study the Kramers-Wannier duality experimentally on a quantum computer. 
The gates appearing in our operator expressions are all two-qubit gates so it should be possible to implement the duality following our quantum operations, say, on an IBM quantum computer using qiskit.
Using our explicitly laid out protocol the transformation of ground states, entanglement entropy,  critical coupling etc., can thus be studied experimentally.

\bibliography{KW}

\end{document}